\documentclass[10 pt, conference]{IEEEtran} 
\usepackage[left=1.62cm,right=1.62cm,top=1.8cm,bottom=2.5cm]{geometry}

\usepackage[T1]{fontenc}
\usepackage{amssymb}
\usepackage{amsfonts}
\usepackage{graphicx}

\usepackage{array}
\usepackage{amsmath}

\usepackage{xcolor}
\usepackage{colortbl}
\usepackage{setspace} 

\usepackage{xcolor,colortbl}
\definecolor{LightCyan}{rgb}{0.88,1,1}
\definecolor{Gray}{gray}{0.85}

\usepackage{graphicx}
\usepackage[outdir=./]{epstopdf}
\usepackage{caption}
\newtheorem{theorem}{Theorem}
\newtheorem{corollary}{Corollary}
\newtheorem{lemma}{Lemma}
\newtheorem{remark}{Remark}

\newtheorem{proposition}{Proposition}

\let\origtau\tau 
\renewcommand{\tau}{\scalebox{1.44}{$\origtau$}}
\allowdisplaybreaks
\usepackage{flushend} 
\title{Stochastic Geometry Analysis of a New GSCM with Dual Visibility Regions}
\date{\today}
\IEEEaftertitletext{\vspace{-1.5\baselineskip}}
\author{
\IEEEauthorblockN{Anish Pradhan\IEEEauthorrefmark{1}, Harpreet S. Dhillon\IEEEauthorrefmark{1}, Fredrik Tufvesson\IEEEauthorrefmark{2}, and Andreas F. Molisch\IEEEauthorrefmark{3}}
\IEEEauthorblockA{\IEEEauthorrefmark{1}The Bradley Department of Electrical and Computer Engineering, Virginia Tech, Blacksburg, USA
\\ \IEEEauthorrefmark{2}Department of Electrical and Information Technology, Lund University, Lund, Sweden \\ \IEEEauthorrefmark{3}Ming Hsieh Department of Electrical and Computer Engineering, University of Southern California, Los Angeles, USA \\
Email:\{pradhananish1, hdhillon\}@vt.edu, fredrik.tufvesson@eit.lth.se, molisch@usc.edu. \vspace{-7 mm}}
\thanks{This work was supported by U.S. National Science Foundation under Grant ECCS-2030215, CNS-2225511, CNS-2107276, and CNS-2106602.
}              
}
\IEEEoverridecommandlockouts
\begin{document}
\maketitle
\begin{abstract}
The geometry-based stochastic channel models (GSCM), which can describe realistic channel impulse responses, often rely on the existence of both {\em local} and {\em far} scatterers.  However, their visibility from both the base station (BS) and mobile station (MS) depends on their relative heights and positions. For example, the condition of visibility of a scatterer from the perspective of a BS is different from that of an MS and depends on the height of the scatterer. To capture this, we propose a novel GSCM where each scatterer has dual disk visibility regions (VRs) centered on itself for both BS and MS, with their radii being our model parameters. Our model consists of {\em short} and {\em tall} scatterers, which are both modeled using independent inhomogeneous Poisson point processes (IPPPs) having distinct dual VRs. We also introduce a probability parameter to account for the varying visibility of tall scatterers from different MSs, effectively emulating their noncontiguous VRs. Using stochastic geometry, we derive the probability mass function (PMF) of the number of multipath components (MPCs), the marginal and joint distance distributions for an active scatterer, the mean time of arrival (ToA), and the mean received power through non-line-of-sight (NLoS) paths for our proposed model. By selecting appropriate model parameters, the propagation characteristics of our GSCM are demonstrated to closely emulate those of the COST-259 model.
\end{abstract}
\section{Introduction}
Owing to their physics-inspired construction and measurement-driven accuracy, GSCMs have become an integral part of the design and analysis of many wireless communications systems. These models place scatterers at random in a geometric plane to obtain the directionally resolved impulse response or transfer function from simplified ray tracing. {There is also a different (though some relations exist) type of model sometimes called GSCM, namely the 3GPP-type channel models where delays and directions of rays are selected at random \cite[Sec. 7.5-7.6]{molisch2023wireless}.}  Typically, GSCMs consider two types of scatterers: a) \textit{local scatterers}, that lie close to the MS and often represent scattering from cars, trees, and buildings in the vicinity of the MS, and b) \textit{far scatterers}, which lie farther from both the BS and MS and often represent far-away high-rise buildings (in urban environments) or mountains (in rural environments). Since point processes are often used to describe the spatial distribution of these scatterers, one can hope to study their properties analytically using ideas from stochastic geometry. \textcolor{black}{It is rather surprising that this connection has not been explored in sufficient detail yet, perhaps because of the lack of interaction between the propagation and stochastic geometry communities. The main objective of this paper is to fill this gap by introducing a novel GSCM that adapts to changes in both the height and horizontal position of BSs, thus altering the visibility of different types of scatterers. This approach allows us to consider two distinct types of scatterers, \textit{tall} and \textit{short}, each with dual visibility regions, making it particularly suited for stochastic geometry analysis.}

\subsection{Prior work} 
Given the topic of this paper, the following two directions of research are the most relevant to this discussion: (i) design of GSCMs in the propagation community, and (ii) stochastic geometry analysis of simple setups in which both users and environmental scatterers/obstacles are stochastically modeled.

Starting with the first direction, GSCMs are well-studied in the literature, and a plethora of analytical results exist for different spatial distributions of the scatterers. The first GSCM, placing scatterers on a circle, was introduced in \cite{Lee}. Subsequent models explored different scatterer distributions such as ellipses \cite{lib}, with \cite{Blanz} extending this to three dimensions. Further models considered multiple circular areas \cite{Norklit} and additional far scatterers \cite{fuhl}. In all these papers, all modeled scatterers are assumed to be visible by both BS and MS. The COST family of GSCMs \cite{cost1,cost2} introduced the concept of noncontiguous VRs to account for the experimentally observed fact that far scatterers have an impact on the channel impulse response only when the MS is in certain regions of a cell. In addition, there are local scatterers that lie within a disk centered around the MS. Moreover, the COST IRACON channel model for vehicle-to-vehicle communication introduced scatterer-centric visibility and gain functions, where scatterers are distributed in the environment according to the layout of the intersection \cite{costiracon}. However, such COST models aim to describe realistic channel measurements without considering analytical tractability.

Coming to the second direction, there has been some relatively recent work in the stochastic geometry literature that explicitly models environmental obstacles along with the locations of the wireless nodes. This has mostly been done in the context of 5G millimeter wave (mmWave) system design. The earliest work in this direction is \cite{bai22}, which considered a stochastic model for the environmental blockages in a mmWave system and analyzed the performance of such a system using stochastic geometry. A similar study focusing on the blind spot probability in localization networks was done by the authors of this paper in \cite{corrblocking}. An author of this paper has also studied the effect of single-reflections in such models \cite{clone1,clone2}, which led to the analytical characterization of ToA, angle of arrival (AoA), NLoS bias, as well as the localization performance in such setups. Recently, the authors of \cite{pandey2023coverage} derive the coverage probability and exact mean signal-to-noise-plus-interference ratio of a terahertz cellular network considering scatterers. While the first-order models studied in this line of work are reasonable, they were not necessarily intended to describe the propagation characteristics of a GSCM.

Therefore, while these two directions are clearly related, the loop was never closed between them, which is the main inspiration behind this paper. 

\subsection{Contributions}
The key contribution of this work is to bridge the gap between the channel modeling works related to GSCM in the propagation community and some related stochastic geometry efforts. In particular, we model the dual VRs of each scatterer for both BS and MS as disks centered at that scatterer. This, along with the modeling of short and tall scatterers as two independent IPPPs and the introduction of a probability parameter to account for the varying visibility of tall scatterers, results in a tractable GSCM. The tractability is demonstrated through the derivation of (i) the distribution of the number of MPCs, (ii) the joint and marginal distance distributions associated with an active scatterer, (iii) an analytical expression of the mean ToA, and (iv) the expected received power through NLoS paths assuming a narrow-band communication and normally distributed reflection or scattering coefficients. Finally, we demonstrate that our GSCM can be fine-tuned to a specific scenario to describe some of the NLoS characteristics of significantly more complicated GSCMs, such as COST-259 for the same scenario. As a concrete case study, we show that our proposed model closely resembles the COST-259 model for the generalized typical urban (GTU) scenario in aspects of the distribution of the number of MPCs, mean ToA, and mean NLoS received power. It also successfully reflects the angular statistics of the scatterers in the COST-259 model to a considerable extent.

\section{Dual Visibility Region-based Scattering Model}
\begin{figure}
 \centering {\includegraphics[clip, trim=2.2cm 5.7cm 14.5cm 3.6cm,width=0.6\columnwidth]{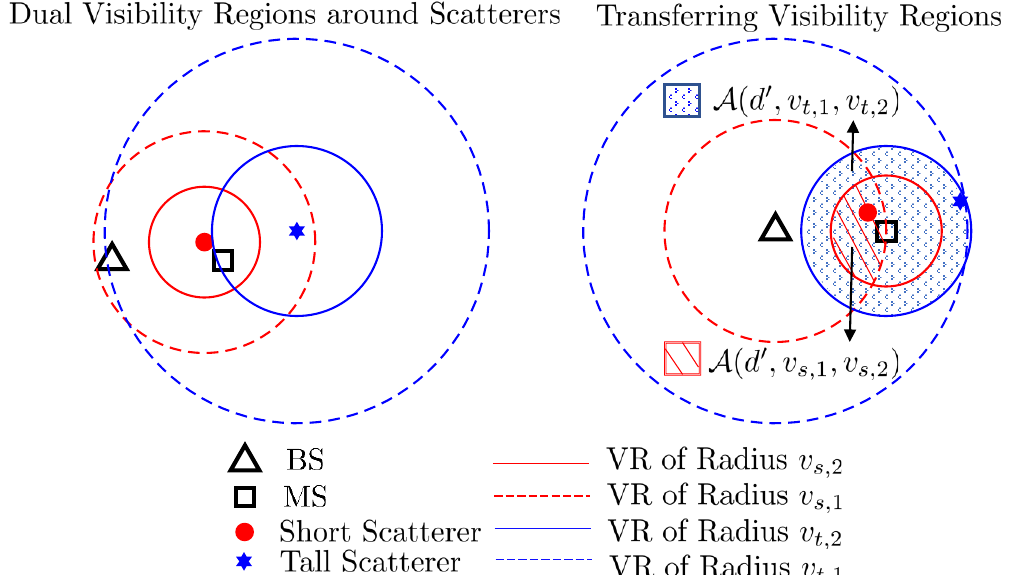}}
        \caption{Illustration of Visibility Regions.}
        \label{fig:sc2}
\end{figure}

Fig. \ref{fig:sc2} demonstrates the VRs along with the BS, the MS, and two active scatterers. The BS is located at the origin $\mathbf{o}=(0,0)$ of the two-dimensional (2D) Cartesian coordinate system and the MS is located at $\mathbf{o'}$ such as $r_{\bf o,o'}=d'$ where $r_{\mathbf{p}_0,\mathbf{y}}=\|\mathbf{p}_0-\mathbf{y}\|$ is the horizontal (2D) distance between a point located at $\mathbf{p}_0$ and a point at $\mathbf{y}$. Additionally, the scatterers are modeled as realizations of point processes similar to \cite{hyperbole,ToA1} and consider two types of independent scatterers, i.e., short and tall which are represented by subscripts `${s}$' and `${t}$', respectively. \textcolor{black}{In particular, the short scatterers are distributed as IPPP ${\Phi}_s$ with constant density $\lambda_s$ while the tall scatterers are distributed as IPPP ${\Phi}_t$ with random density $U\lambda_t$, where $U$ is a Bernoulli random variable with mean $\gamma$. Therefore, the point process of the scatterers actually becomes a Cox process or a doubly stochastic Poisson process (since its intensity is random)~\cite{stoyan}. We will explain the rationale behind including $U$ after introducing the VRs next.}


We define the VRs for the BS and the MS around a scatterer of type $k\in\{s,t\}$ located at ${\mathbf p}_k$ as concentric circles $b({\mathbf p}_k, v_{k, 1})$ and $b({\mathbf p}_k, v_{k, 2})$, respectively, where $b({\mathbf p}, r)$ denotes a circle of radius $r$ centered at ${\mathbf p}$. While this is meaningful for short scatterers, we will justify this for tall scatterers shortly. The BS or MS is said to be in the LoS of a given scatterer if it lies within the respective VR of that scatterer. Note that the active scatterers that contribute to the MPCs must exist and be visible to both the BS and the MS. In other words, for an existing scatterer to be active, both the BS and the MS must be inside their respective VRs centered on that scatterer. We also assume that each MPC from the BS to the MS interacts with a single active scatterer. Consequently, the active scatterers follow an inhomogeneous PPP with density
\textcolor{black}{
\begin{align}
\lambda_{k,v}=
\begin{cases}
   \lambda_s, &k=s,~ r_{\mathbf{p}_s,\mathbf{o}}\leq v_{s,1}, ~ r_{\mathbf{p}_s,\mathbf{o^\prime}}\leq v_{s,2} \\
   U\lambda_t, &k=t,~ r_{\mathbf{p}_t,\mathbf{o}}\leq v_{t,1}, ~ r_{\mathbf{p}_t,\mathbf{o^\prime}}\leq v_{t,2} \\
   0,  &\text{otherwise}
\end{cases}. \label{eq:inhomPPP}
\end{align}}
\begin{remark} \label{rem:vrtransfer}
\textcolor{black}{Considering a circular VR $b({\mathbf p}_k, v_{k, 1})$ around the scatterer for the BS, the condition for visibility is $r_{\mathbf{p}_k,\mathbf{o}}\leq v_{k,1}$. Similarly, drawing the VR around the BS results in the same condition for visibility $r_{\mathbf{o},\mathbf{p}_k}\leq v_{k,1}$. This implies that a circular VR around a scatterer of type $k$ for either BS or MS is equivalent to having that VR around that BS or MS. As a direct consequence, we can observe that transferring the VRs around the BS and MS as their centers and denoting the active scatterers in their intersection area is equivalent to having both VRs around the scatterers and choosing active scatterers from (\ref{eq:inhomPPP}). For example, if we draw a VR of radius $v_{k,1}$ around the BS and another VR of radius $v_{k,2}$ around the MS, the scatterers of type $k$ inside the intersection area will follow (\ref{eq:inhomPPP}). Based on this observation, Fig. \ref{fig:sc2} illustrates the transferring of the VRs around the BS and MS.}
\end{remark}

In our model, the VR's radius depends on the scatterer's height, and the height of the BS or the MS. Following this approach, the probability of visibility of a scatterer from a node depends on the distance between them. Note that the concept of short and tall scatterers corresponds to the idea of local and far scatterers in the relevant literature \cite{fuhl,cluster}. Interestingly, choosing $v_{s,1}\gg v_{s,2}$ gives rise to a local cluster of scatterers around MS which imitate a generic macrocell model where MPCs are less likely around the BS. \textcolor{black}{The height of scatterers primarily impacts the VR radius, while the effect on elevation angles is disregarded due to the 2D coordinate system analysis.}

\textcolor{black}{Now, we are ready to explain the reason for introducing the random variable $U$ in the density of the tall scatterers. As discussed above already, this basically means that the tall scatterer will be visible if two conditions are satisfied: (i) the value of $U$ is 1 (which happens with probability $\gamma$), and (ii) the condition for visibility described in Remark~\ref{rem:vrtransfer} is satisfied.} Note that the VRs of tall scatterers are usually noncontiguous. Therefore, some MSs will see the tall scatterers while others would not. Despite considering a simple circular VR around the tall scatterers, we are able to emulate this effect through parameter $\gamma$. We will demonstrate this through a comprehensive case study of COST-259 in Section~\ref{sec:sim}. However, it is crucial to exercise caution since this approach is only adequate for examining the channel's {\em marginal} properties (at a single time and location). To investigate joint properties such as correlations across time and space, we will need to extend this model, which presents a promising direction for future research. 

\textcolor{black}{ Note that in our GSCM implementation, the superposition of different MPCs inherently provides small-scale fading. However, since our primary focus in this paper is on the spatial aspects of multipath propagation, we have not included large-scale fading (shadowing) in this study. One way of including large-scale fading in our model is through {\em cluster shadow fading} using separate stochastic processes for each cluster or group of scatterers with similar delay \cite{cost2}. 
Since these processes might exhibit correlation, their inclusion in our model presents an intriguing opportunity for future research.}
\section{Mathematical Analysis} \label{sec:TA}
In this section, we derive the PMF of the number of MPCs, joint distance distributions, mean ToA, and the mean received power through NLoS paths for our proposed model. These results have useful applications in various wireless applications such as synchronization and localization. First, as noted in Remark \ref{rem:vrtransfer}, the intersection area between two circles plays a significant role in the analyses. For concise notation in later derivations, we define $\mathcal{A}(d_0,a,b)$, as depicted in Fig. \ref{fig:sc2}, which denotes the intersection area of two circles of radii $a$ and $b$ whose centers are separated by distance $d_0$ as:
\begin{equation}
    \mathcal{A}(d_0,a,b)=
    \begin{cases}
        0, & a+b< d_0 \\
       \pi \min(a,b)^2, & |a-b|> d_0 \\
       \mathcal{A}'(d_0,a,b), & \text{otherwise}
    \end{cases}.
\end{equation}
where, $\mathcal{A}'(d_0,a,b)$ is defined below in \eqref{eq:area}.
\begin{align}
  &\mathcal{A}'(d_0,a,b)=\notag b^2\cos\!^{-1}\left(\frac{d_0^2\!+\!b^2-a^2}{2d_0b}\right)+\\& a^2\cos\!^{-1}\left(\frac{d_0^2\!+\!a^2-b^2}{2d_0a}\right)-\frac{1}{2}\sqrt{4d_0^2a^2-(d_0^2-b^2\!+\!a^2)^2}. \label{eq:area}
\end{align}

We begin our analysis by characterizing the number of MPCs. In Lemma \ref{lem:MPCd}, we show that the PMF of the number of the MPCs is a Poisson distribution following directly from the PPP assumption in our model.
\begin{lemma}\label{lem:MPCd}
The number of MPCs $N$ has a PMF $f_N(n)$ with mean $\mu=\mu_s+\gamma\mu_t$, where $\mu_k=\lambda_k\mathcal{A}(d',v_{k,1},v_{k,2}) \forall k\in\{s,t\}$. $f_N(n)$ is described below:
\begin{align}
	f_N(n)=\gamma P(n,\mu_s+\mu_t)+(1-\gamma) P(n,\mu_s),
\end{align}
where $P(m,\varphi)=\frac{e^{-\varphi}\varphi^m}{m!}$ denotes a generic Poisson PMF.
\end{lemma}
\begin{IEEEproof}
See Appendix \ref{sec:proofMPC}.
\end{IEEEproof}
\textcolor{black}{The above Lemma, which is a direct consequence of the PPP assumption on scatterers, is very useful in later derivations of both the mean ToA and mean received power. Next, we aim to determine the mean ToA in Proposition \ref{prop:meanTOA}. To do so, we first derive the marginal cumulative distribution functions (CDF) of the distances traveled by MPCs in Lemma \ref{lem:distcdf}, which is described below.}
\begin{lemma}
 If an active scatterer of type $k$ is situated at $\mathbf{p}_k$, then $X_k=r_{\mathbf{p}_k,\mathbf{o}}$ and $Y_k=r_{\mathbf{p}_k,\mathbf{o'}}$ denote the distance RVs. The CDFs of $X_k$ and $Y_k$ are given as:
\begin{align}
F_{X_k}(x_k)=
\frac{\mathcal{A}(d',x_k,v_{k,2})}{\mathcal{A}(d',v_{k,1},v_{k,2})}, \quad a_{\min}<x_k<a_{\max},
\end{align}
where $a_{\min}=\max(d'-v_{k,2},0)$ and $a_{\max}=\min(d'+v_{k,2},v_{k,1})$.
\begin{align}
F_{Y_k}(y_k)=
\frac{\mathcal{A}(d',y_k,v_{k,1})}{\mathcal{A}(d',v_{k,1},v_{k,2})}, \quad b_{\min}<x_k<b_{\max},
\end{align}
where $b_{\min}=\max(d'-v_{k,1},0)$ and $b_{\max}=\min(d'+v_{k,1},v_{k,2})$.
\label{lem:distcdf}
\end{lemma}
\begin{IEEEproof}
See Appendix \ref{sec:proofdistcdf}.
\end{IEEEproof}
\textcolor{black}{Following these marginal distance CDFs, the mean ToA $\mu_{\rm ToA}$ is defined as
\begin{align}
    \mu_{\rm ToA}=\frac{\mathrm{E}\left[\tau\right]}{c},
\end{align}
where $\tau=X_k+Y_k$  denotes the distance covered by an MPC, and $c$ is the speed of light.
As the CDFs have positive support, we present the expected values of these RVs in the following corollary to facilitate the derivation of $\mu_{\rm ToA}$.}
\begin{corollary}
The mean distance between the BS and an active scatterer of type $k$ is $\mathrm{E}[X_k]=\int\limits_{0}^{a_{\max}}(1-F_{X_k}(x_k))\mathrm{d} x_k$, and the mean distance between an active scatterer of type $k$ is $\mathrm{E}[Y_k]=\int\limits_{0}^{b_{\max}}(1-F_{Y_k}(y_k))\mathrm{d} y_k$. \label{corr:mean}
\end{corollary}
Using the expected values of $X_k$ and $Y_k$, we finally derive the mean ToA in Proposition \ref{prop:meanTOA} next.
\begin{proposition} \label{prop:meanTOA} 
The mean ToA in our proposed model is
\begin{align}
 \mu_{\rm ToA}=
   \frac{\mathrm{E}\left[\tau\right]}{c}, 
\end{align}
where $\mathrm{E}\left[\tau\right]=\gamma\sum\limits_{k\in\{s,t\}}\frac{\mu_k}{\mu}\left(\mathrm{E}[X_k]+\mathrm{E}[Y_k]\right)+(1-\gamma)\left(\mathrm{E}[X_s]+\mathrm{E}[Y_s]\right).$
\end{proposition}
\begin{IEEEproof}
See Appendix \ref{sec:proofToA}.
\end{IEEEproof}

Now, we present the NLoS mean received power using the \textcolor{black}{ray-tracing models} in \eqref{eq:rxpower} considering that all the MPCs are going through the same electromagnetic phenomena but with different parameters \cite{goldsmith_pathloss}. Note that this equation is based on the assumptions of a narrowband channel and isotropic antenna with unit gain.
\begin{align}
P_r =& k_0 \bigg|\sum\limits_{k\in\{s,t\}}\sum_{i=1}^{N_k}\frac{R_k^i\exp\!\bigg(-j\theta(X_k^i,Y_k^i)\bigg)}{g_2(X_k^i,Y_k^i)}\bigg|^2, \label{eq:rxpower}
\end{align}
where $P_r$ is the received power, $k_0$ is the transmission constant, $\lambda$ is the wavelength, $N_k$ is the number of active scatterers of type $k$, $R_k^i$ is the reflection or scattering coefficient RV for the $i$-th active scatterer of type $k$, $\theta(X,Y)=\frac{2\pi g_1(X,Y)}{\lambda}$, $g_1(X,Y)$, $g_2(X,Y)$ are functions of $X$ and $Y$, $X_k^i$ is the distance between the $i$-th active scatterer of type $k$ and the BS, and $Y_k^i$ is the distance between the $i$-th active scatterer of type $k$ and the MS through the $i$-th scatterer. Note that the choice of $k_0$, $g_1(X,Y)$, $g_2(X,Y)$, and $R_k^i$ dictates how the signal electromagnetically interacts with the scatterers. For example, in the case of scattering, $g_1(X,Y)=X+Y$, $g_2(X,Y)=XY$, and $k_0=P_t\frac{\lambda^2}{(4\pi)^{3}}$ and in the case of reflection, $g_2(X,Y)=X+Y=g_1(X,Y)$ and $k_0=P_t\left(\frac{\lambda}{4\pi}\right)^2$ \cite{goldsmith_pathloss}, where $P_t$ denotes the transmit power. However, calculating $\mathrm{E}[P_r]$ requires the joint distance distribution of $X_k$ and $Y_k$ for each of the active scatterers, which is presented in Lemma \ref{lem:jpdf}.
\begin{lemma}
The joint probability distribution function (PDF) of $X_k$ and $Y_k$ for an active scatterer is:
\begin{align}
    f_{X_k,Y_k}(x_k,y_k)=\begin{cases}
       \frac{\mathcal{B}(d',x_{k},y_{k})}{\mathcal{A}(d',v_{k,1},v_{k,2})}, & \begin{aligned}
       &x_{\min}<x_k<x_{\max} \\ &y_{\min}<y_k<y_{\max}  
       \end{aligned}\\
       0, & \text{otherwise}
    \end{cases}, 
\end{align}
where, $\mathcal{B}(d',x_{k},y_{k})$ is defined in \eqref{eq:jpdf}, $x_{\min}=\max(d'-v_{k,2},0)$, $x_{\max}=\min(d'+v_{k,2}, v_{k,1})$, $y_{\min}=\max(d'-x_k,0)$, and $y_{\max}$ is described in Table \ref{tab:ymax}. \label{lem:jpdf}
\end{lemma}
\begin{IEEEproof}
See Appendix \ref{sec:proofjpdf}.
\end{IEEEproof}
\setlength{\tabcolsep}{3.5pt}
\begin{table}
    \centering
\begin{tabular}{|c|c|}
\hline
    Cases & $y_{\max}$ \\ \hline
    $v_{k,1}-v_{k,2}\geq d'$ &  $v_{k,2}$ \\ \hline
    $v_{k,2}-v_{k,1}\geq d'$ &  $\min(d'+x,d'+v_{k,1})$ \\ \hline
    $|v_{k,1}-v_{k,2}|<d',v_{k,1}\text{ \& } v_{k,2}<d'$ &  $v_{k,2}$ \\ \hline
    $|v_{k,1}-v_{k,2}|<d',v_{k,1} \text{ or } v_{k,2}>d'$ &  $\min(d'+x,v_{k,2})$ \\ \hline
\end{tabular}
    \caption{$y_{\max}$ in different parameter combinations.}
    \label{tab:ymax}
\end{table}
\begin{figure*}
\begin{align}
    &\mathcal{B}(d',x_{k},y_{k})=\frac{x_k}{d'\,\sqrt{1-\frac{{\left(d'^2-{x_k}^2+{y_k}^2\right)}^2}{4\,d'^2\,{y_k}^2}}}+\frac{4\,x_k\,y_k^3\,\left(d'^2+{x_k}^2-{y_k}^2\right)}{{\left(4\,d'^2\,{x_k}^2-{\left(d'^2+{x_k}^2-{y_k}^2\right)}^2\right)}^{3/2}}-\frac{x_k\,\left(d'^4-2\,d'^2\,{x_k}^2+{x_k}^4-{y_k}^4\right)}{4\,d'^3\,{y_k}^2\,{\left(1-\frac{{\left(d'^2-{x_k}^2+{y_k}^2\right)}^2}{4\,d'^2\,{y_k}^2}\right)}^{3/2}}\label{eq:jpdf}
\end{align}
\hrule
\vspace{-2 mm}
\end{figure*}
Armed with this joint distribution, mean received power numerically can now be found by using the joint PDF described in Lemma \ref{lem:jpdf} and \eqref{eq:rxpower} through the Theorem \ref{theo:meanRx}, which is presented next.
\begin{theorem}
The mean received power is:
\begin{align}
    &\mathrm{E}[P_r]=\gamma k_0\bigg(\sum\limits_k\left((g_k+h_k^2+g'_k+h_k^{\prime^2})\mu_k\right)+\\\notag
    &(\sum\limits_k h_k\mu_k)^2+(\sum\limits_k h'_k\mu_k)^2\bigg)+ (1-\gamma) k_0\bigg((g_s+h_s^2+g'_s+\\ \notag
    &h_s^{\prime^2})\mu_s+( h_s\mu_s)^2+( h'_s\mu_s)^2\bigg),
\end{align}
where, 
\begin{align*}
&h_k\!=\!\mathrm{E}\!\Bigg[\frac{\!R_k^i\!\cos\!\bigg(\!\theta(X_k^i,Y_k^i)\!\bigg)}{g_2(X_k^i,Y_k^i)}\!\Bigg],g_k\!=\!\mathrm{Var}\!\Bigg[\frac{\!R_k^i\!\cos\!\bigg(\!\theta(X_k^i,Y_k^i)\!\bigg)}{g_2(X_k^i,Y_k^i)}\!\Bigg],
\\&h'_k\!=\!\mathrm{E}\!\Bigg[\frac{\!R_k^i\!\sin\!\bigg(\!\theta(X_k^i,Y_k^i)\!\bigg)}{g_2(X_k^i,Y_k^i)}\!\Bigg], g'_k\!=\!\mathrm{Var}\!\Bigg[\frac{\!R_k^i\!\sin\!\bigg(\!\theta(X_k^i,Y_k^i)\!\bigg)}{g_2(X_k^i,Y_k^i)}\!\Bigg].
\end{align*}
 \label{theo:meanRx}
\end{theorem}
\begin{IEEEproof}
See Appendix \ref{sec:proofmeanrx}.
\end{IEEEproof}

Although we demonstrated the analytical tractability of our proposed model with a comprehensive theoretical analysis, we have yet to show its applicability. To do so, we compare its scattering characteristics with the COST model in a comprehensive case study presented in the next section.

\section{Simulation Results} \label{sec:sim}
\begin{figure*}
\minipage{0.25\textwidth}
   \centering    \includegraphics[width=\columnwidth]{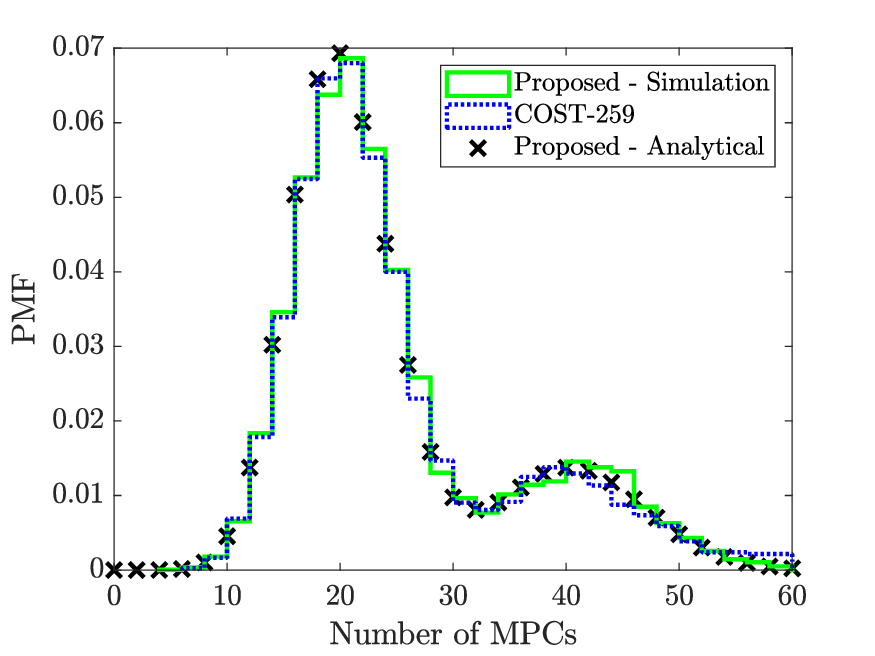}
    \caption{MPC PMF $f_N(n)$.}
    \label{fig:mpc}
    \endminipage\hfill
\minipage{0.25\textwidth}
    \centering     \includegraphics[width=\columnwidth]{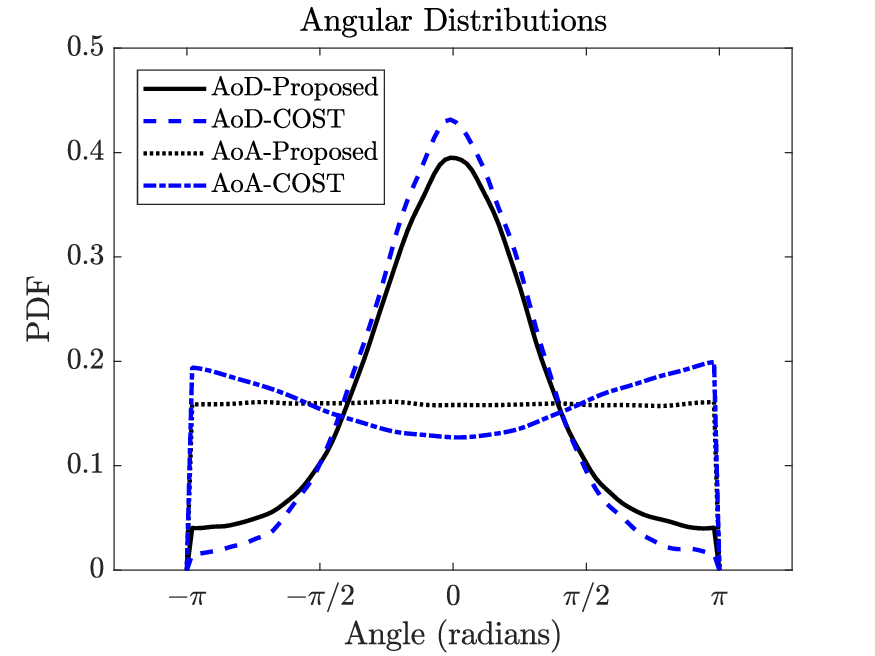}
    \caption{AoD and AoA Statistics.}
    \label{fig:angle}
    \endminipage\hfill
    \minipage{0.25\textwidth}
    \centering     
    \includegraphics[width=\columnwidth]{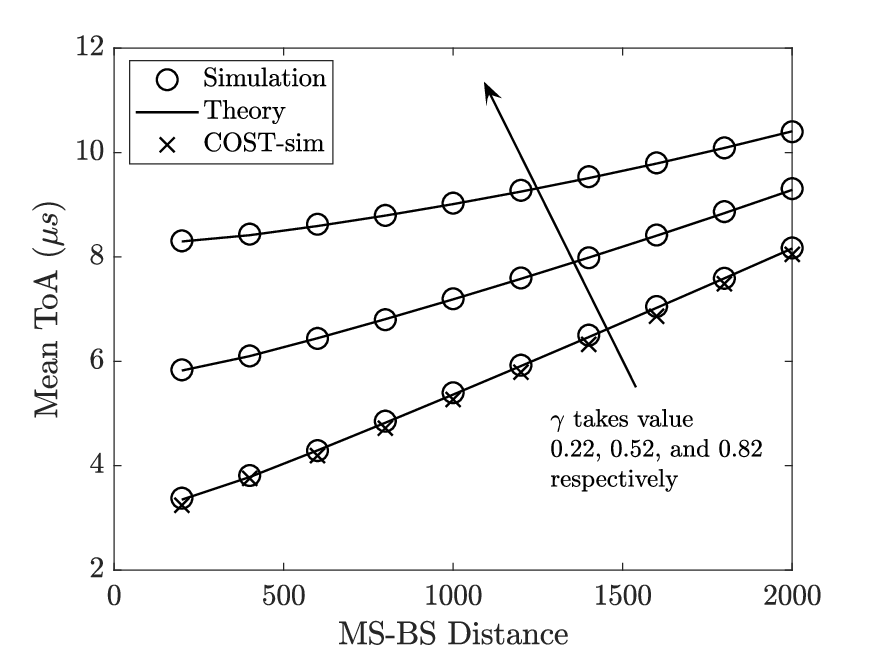}
    \caption{Mean ToA v/s $d'$.}
    \label{fig:meandist}
    \endminipage\hfill
     \minipage{0.25\textwidth}
    \centering     
    \includegraphics[width=\columnwidth]{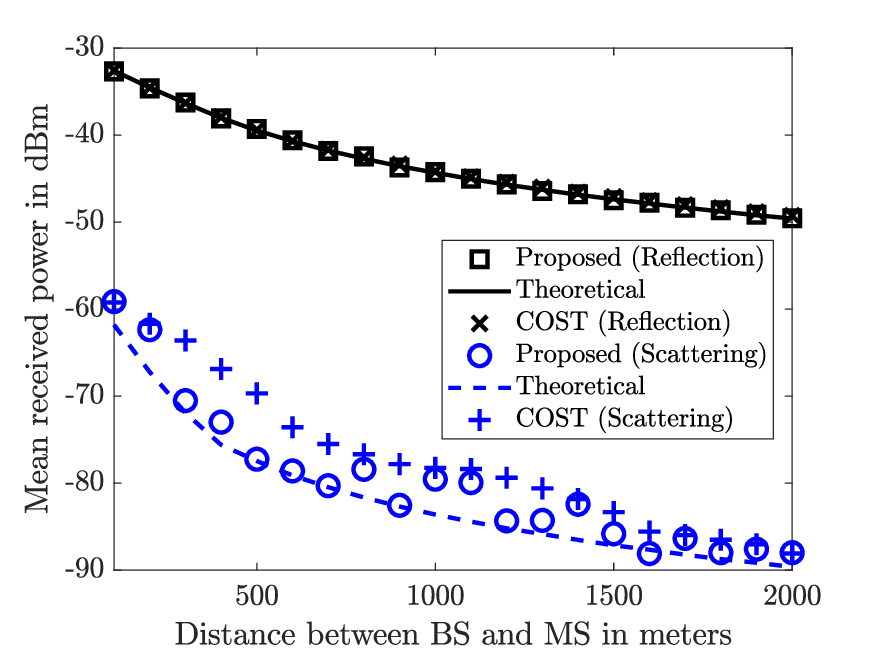}
    \caption{Mean received power.}
    \label{fig:rxpowerfig}
    \endminipage\hfill
\end{figure*}


In this section, we compare the propagation characteristics of our proposed GSCM with COST-259. For comparison, a GTU scenario with a cell radius of $3$ kilometer is chosen for COST-259 considering only 2D angular measurements. 

The scatterers are distributed as PPPs in disks of radii $300$ meter \cite{fuhl,cluster} centered on cluster positions. The intensities of the scatterers in the local cluster and far clusters are chosen in a way such that the mean number of scatterers in each cluster is $20$. \textcolor{black}{The GTU scenario is simulated with the COST-259 model by using the default parameters provided by \cite{cluster}.}
\begin{table}
    \centering
    \begin{tabular}{|c|c|c|c|c|c|c|c|c|}
    \hline 
        Case & $d'$ & $v_{t,1}$ & $v_{t,2}$ & $v_{s,1}$ & $v_{s,2}$ & $\frac{\lambda_s}{10^{-5}}$ & $\frac{\lambda_t}{10^{-7}}$ & $\gamma$ \\ \hline
        GTU & $0.2$ & $4.1$ & $4$ & $0.5$ & $0.3$ & $7.07 $ & $4.2$ & $0.22$ \\ \hline
    \end{tabular}
    \caption{GSCM parameters.}
    \label{tab:case}
\end{table}
Furthermore, we simulate the GTU scenario in our GSCM according to the parameters in Table \ref{tab:case}. In this scenario, we choose $v_{s,1}$ such that $b(\mathbf{o},v_{s,1})$ always encompasses $b(\mathbf{o'},v_{s,2})$ to preserve the corresponding local cluster characteristics in COST-259. \textcolor{black}{The radii parameters for the tall scatterers are chosen via inspection while a better method will be investigated in future works.} For both of the models, the MS is situated at $\mathbf{o'}=(d',0)$ with $d'=0.2$km. All the distance parameters are in kilometers. For the mean received power simulation, we use $10$ W transmit power, $2$ GHz operating frequency, and normally distributed $R_i^k$ with arbitrary mean $\mu_R$ and variance $\sigma_R^2$. For reflection, $\mu_R=-1.17$, $\sigma_R^2=0.4$, and for scattering, $\mu_R=4$, $\sigma_R^2=2$ are used. These parameters are selected to obtain realistic coefficients per \cite{goldsmith_pathloss}. Note that we consider all the active scatterers in both models for a fair comparison.

We plot the PMF of the number of MPCs in Fig. \ref{fig:mpc} for our proposed model and for COST-259 in the GTU scenario to demonstrate the performance of our proposed model. The COST-259 PMF has twin peaks where each peak corresponds to the effect of the local scatterers and the compound effect of the far scatterers. This happens as local scatterers are more common while far scatterers are rare in a GTU scenario \cite{cost2}. The first peak denotes a higher probability of the existence of only local scatterers, whereas the second peak denotes a lower probability of both local and far scatterers existing simultaneously. This can be perfectly captured by our proposed GSCM with appropriate parameters which are chosen to replicate the GTU scenario. We also demonstrate that the analytical PMF derived in Lemma \ref{lem:MPCd} is exact.

We plot the AoA and angle of departure (AoD) PDFs for the GTU scenario in Fig. \ref{fig:angle} to compare the angular statistics between our proposed model and the COST model. We can observe that the AoD distribution is quite similar and centered at zero for both models as the MS is situated on the positive x-axis. However, the AoA distribution for our proposed model is almost uniform while the AoA distribution of the COST model is slightly distorted from the uniform shape. \textcolor{black}{In our proposed model, the radii and intensity parameters regulate a localized scattering disk and rare tall scatterers spread over a wide zone without any angular bias, leading to a uniformly shaped AoA distribution.} In contrast, as the COST model favors the far scatterers which are more aligned with the BS and the MS, many far scatterers are generated on the side of the BS resulting in higher probabilities around $\pi$ and $-\pi$ in the distribution. However, we demonstrate that our proposed model captures the angular statistics of the COST model to a good extent.

We plot the mean ToA in microseconds with the distance between MS and BS $d'$ in Fig. \ref{fig:meandist}. Curves with different $\gamma$ are plotted to demonstrate the effect of increasing $\gamma$ on mean ToA. We can observe an increasing mean ToA trend with increasing $\gamma$ due to the increase of the probability of the existence of tall scatterers. As the curve corresponding to $\gamma=0.22$ denotes the GTU scenario, we also plot the COST-259 simulation results. We can observe that our proposed GSCM is able to imitate the mean ToA achieved by the COST model.

In Fig. \ref{fig:rxpowerfig}, we plot the mean received power through the NLoS paths for our proposed GSCM and COST-259 model. The black curve denotes the case when the electromagnetic interaction is solely a reflection while the black curve denotes the case of only scattering. In general, the actual interaction consists of a combination of the two. \textcolor{black}{Note that the fluctuations for the scattering case simulations can be attributed to the inaccuracy in the numerical integration procedures.} However, the accuracy of our proposed GSCM against COST in both those extreme scenarios implies its applicability to capture more realistic electromagnetic interactions as well.

\section{Conclusions}
In this work, we introduced a novel GSCM that accounts for two distinct types of scatterers with dual VRs, making it more amenable to stochastic geometry analysis. The scatterers were modeled as independent IPPPs, and the VRs are represented as concentric circles centered at the scatterers. We also incorporated a probability parameter to emulate the varying visibility of tall scatterers from different MSs. Our GSCM's tractability was demonstrated by deriving the PMF of the number of MPCs, joint and marginal distance distributions associated with an active scatterer, mean ToA, and mean received power through NLoS paths. Furthermore, as a part of a concrete case study, we showed that our GSCM can be tailored to specific scenarios, such as the GTU, to capture some of the NLoS propagation characteristics of the COST-259 model for the same scenario. In particular, our proposed model accurately describes the distribution of the number of MPCs, mean ToA, and mean received power of the COST-259 model for the GTU scenario. Additionally, it captures the scatterers' angular statistics in the COST-259 model to a significant degree. There are many avenues for future work. Specifically, this involves expanding scatterer modeling with a Gaussian intensity function, extending the GSCM to spatially and temporally consistent models, incorporating fading characteristics, \textcolor{black}{utilizing measurement data to estimate model parameters}, and analyzing power delay profiles.
\appendix
\subsection{Proof of Lemma \ref{lem:MPCd}:} \label{sec:proofMPC}
The number of MPCs created by scatterers of type $k$ is equal to the number of active scatterers of that type as each signal is assumed to interact with a single active scatterer. Using Remark \ref{rem:vrtransfer} and \eqref{eq:inhomPPP}, $N_k$ active scatterers are distributed as a PPP in the intersection area of the VRs $b(\mathbf{o},v_{k,1})$ and $b(\mathbf{o'},v_{k,2})$ around BS and MS respectively. By the properties of PPP, $N_k\sim\mathrm{Pois}(\mu_k)$, which denotes that $N_k$ is a Poisson RV with mean $\mu_k$ where $\mu_k=\lambda_k\mathcal{A}(d',v_{k,1},v_{k,2})$. Then, the number of all active scatterers is $N=N_s+UN_t$, where $U$ is a Bernoulli RV with mean $\gamma$. The PMF and its mean are derived by deriving the CDF of $N$ using the law of total probability.
\subsection{Proof of Lemma \ref{lem:distcdf}:} \label{sec:proofdistcdf}
$F_{X_k}(x_k)$ is defined in terms of probability below.
\begin{align}
F_{X_k}(x_k)=\mathrm{Pr}[X_k\leq x_k]=\mathrm{Pr}[r_{\mathbf{p}_k,\mathbf{o}}\leq x_k].
\label{eq:lem3proof}
\end{align}
As $r_{\mathbf{p}_k,\mathbf{o'}}\leq v_{k,2}$, \eqref{eq:lem3proof} denotes the probability that an active scatterer is situated in the intersection of $b(\mathbf{o},x_k)$ around BS and $b(\mathbf{o'},v_{k,2})$ around MS. The resulting intersection area on which the active scatterer must lie is $\mathcal{A}(d',x_k,v_{k,2})$. As the active scatterers are uniformly distributed in $\mathcal{A}(d',v_{k,1},v_{k,2})$,
\begin{align}
F_{X_k}(x_k)=\mathrm{Pr}[r_{\mathbf{p}_k,\mathbf{o}}\leq x_k]=\frac{\mathcal{A}(d',x_k,v_{k,2})}{\mathcal{A}(d',v_{k,1},v_{k,2})}.  
\end{align}
 The limits $a_{\min}$ and $a_{\max}$ are found from the intersection points between two circles. Similarly, $F_{Y_k}(y_k)$ is derived.
\subsection{Proof of Proposition \ref{prop:meanTOA}:}\label{sec:proofToA}
By the law of total expectation,
\begin{align*}
    \mathrm{E}[\tau]=\mathrm{E}[\tau|U=1]\mathrm{Pr}[U=1]+\mathrm{E}[\tau|U=0]\mathrm{Pr}[U=0].
\end{align*}
Note that, the second term corresponds to the mean distance covered by the short scatterers multiplied by $1-\gamma$. We derive the first term as:
$\mathrm{E}[\tau|U=1]=\mathrm{E}[X_t+Y_t]\mathrm{Pr}[k=t]+\mathrm{E}[X_s+Y_s]\mathrm{Pr}[k=s]$, where $\mathrm{Pr}[k=t]$ and $\mathrm{Pr}[k=s]$ denote the probabilities of an active scatterer being tall and short respectively. By the properties of Poisson distribution,  $\mathrm{Pr}[k=t]$ and $\mathrm{Pr}[k=s]$ can be shown to be equal to $\frac{\mu_t}{\mu_t+\mu_s}$ and $\frac{\mu_s}{\mu_t+\mu_s}$ respectively. The proposition is proved by replacing these probability values and the expected values calculated from Corollary \ref{corr:mean}. 
\subsection{Proof of Lemma \ref{lem:jpdf}:} \label{sec:proofjpdf}
Following the proof of Lemma \ref{lem:distcdf}, the joint CDF $F_{X_k,Y_k}(x_k,y_k)$ denotes the probability that an active scatterer of type $k$ is situated in the intersection of $b(\mathbf{o},x_k)$ and $b(\mathbf{o'},y_k)$ around BS and MS respectively. Due to uniform distribution of scatterers, the probability is $F_{X_k,Y_k}(x_k,y_k)=\frac{\mathcal{A}'(d',x_k,y_k)}{\mathcal{A}(d',v_{k,1},v_{k,2})}$ where $x_{\min}<x_k<x_{\max}$ and $y_{\min}<y_k<y_{\max}$. The bounds are found from the intersection points of the circles. Now, the joint PDF is calculated by double differentiating $F_{X_k,Y_k}(x_k,y_k)$:
\begin{align}
    \frac{\partial^2F_{X_k,Y_k}(x_k,y_k)}{\partial x_k\partial y_k}=&\notag\frac{\frac{\partial^2\mathcal{A}'(d',x_k,y_k)}{\partial x_k\partial y_k}}{\mathcal{A}(d',v_{k,1},v_{k,2})}=\frac{\mathcal{B}(d',x_k,y_k)}{\mathcal{A}(d',v_{k,1},v_{k,2})}.
\end{align}
This completes the proof.
\subsection{Proof of Theorem \ref{theo:meanRx}:}\label{sec:proofmeanrx}
Similar to the proof of the Proposition \ref{prop:meanTOA}, we use the law of total probability to separate the case $U=1$ and $U=0$. Next, conditioning on $U=1$ and segregating the real cosine terms and imaginary sine terms in \eqref{eq:rxpower}, we can rewrite the expression as:
\begin{align}
    &P_r|(U=1)=\notag k_0\bigg|\sum\limits_k\sum\limits_{i=1}^{N_k}\frac{R_k^i\cos\!\bigg(\theta(X_k^i,Y_k^i)\bigg)}{g_2(X_k^i,Y_k^i)}+\\&j\sum\limits_k\sum\limits_{i=1}^{N_k}\frac{R_k^i\sin\!\bigg(\theta(X_k^i,Y_k^i)\bigg)}{g_2(X_k^i,Y_k^i)}\bigg|^2=k_0|\alpha+j\beta|^2. \label{eq:fmrx}
\end{align}
Assuming that the central limit theorem (CLT) can be applied when conditioned on $N_k=n_k$, we can express the conditional distribution of $\alpha$ and $\beta$ in the following way:
$\alpha|(N_s=n_s,N_t=n_t) \sim \mathcal{N}(\sum\limits_k h_kn_k,\sum\limits_k g_kn_k),\beta|(N_s=n_s,N_t=n_t) \sim \mathcal{N}(\sum\limits_k h'_kn_k,\sum\limits_k g'_kn_k).
$ The $h_k,g_k,h'_k$, and $g'_k$ terms are defined in the theorem and can be calculated numerically through the joint distance distribution and both the mean and variance of $R_k^i$. Now the mean received power can be written in terms of expectation and variance terms of $\alpha$ and $\beta$:
$\mathrm{E}[P_r|U=1]=k_0(\mathrm{Var}[\alpha]+(\mathrm{E}[\alpha])^2+\mathrm{Var}[\beta]+(\mathrm{E}[\beta])^2).$
We replace the unconditional means and variances through the laws of total expectation and variance. The theorem is proved by removing the conditioning by $U$ on the expectation.
\bibliographystyle{IEEEtran}
\setstretch{0.82}
\bibliography{ref2}

\begin{thebibliography}{10}
\providecommand{\url}[1]{#1}
\csname url@samestyle\endcsname
\providecommand{\newblock}{\relax}
\providecommand{\bibinfo}[2]{#2}
\providecommand{\BIBentrySTDinterwordspacing}{\spaceskip=0pt\relax}
\providecommand{\BIBentryALTinterwordstretchfactor}{4}
\providecommand{\BIBentryALTinterwordspacing}{\spaceskip=\fontdimen2\font plus
\BIBentryALTinterwordstretchfactor\fontdimen3\font minus
  \fontdimen4\font\relax}
\providecommand{\BIBforeignlanguage}[2]{{%
\expandafter\ifx\csname l@#1\endcsname\relax
\typeout{** WARNING: IEEEtran.bst: No hyphenation pattern has been}%
\typeout{** loaded for the language `#1'. Using the pattern for}%
\typeout{** the default language instead.}%
\else
\language=\csname l@#1\endcsname
\fi
#2}}
\providecommand{\BIBdecl}{\relax}
\BIBdecl

\bibitem{molisch2023wireless}
A.~F. Molisch, \emph{{Wireless Communications: From Fundamentals to Beyond 5G,
  3rd ed.}}\hskip 1em plus 0.5em minus 0.4em\relax IEEE Press - Wiley, 2023.

\bibitem{Lee}
W.~Lee, ``{Effects on Correlation Between Two Mobile Radio Base-Station
  Antennas},'' \emph{IEEE Trans. on Commun.}, vol.~21, no.~11, pp. 1214--1224,
  Nov. 1973.

\bibitem{lib}
J.~Liberti and T.~Rappaport, ``{A Geometrically based Model for Line-of-Sight
  Multipath Radio Channels},'' in \emph{Proc., IEEE Veh. Technology Conf.
  (VTC)}, vol.~2, Apr. 1996, pp. 844--848.

\bibitem{Blanz}
J.~Blanz and P.~Jung, ``{A Flexibly Configurable Spatial Model for Mobile Radio
  Channels},'' \emph{IEEE Trans. on Commun.}, vol.~46, no.~3, pp. 367--371,
  Mar. 1998.

\bibitem{Norklit}
O.~Norklit and J.~Andersen, ``{Diffuse Channel Model and Experimental Results
  for Array Antennas in Mobile Environments},'' \emph{IEEE Trans. on Antennas
  and Propagation}, vol.~46, no.~6, pp. 834--840, June 1998.

\bibitem{fuhl}
J.~{Fuhl}, A.~F. {Molisch}, and E.~{Bonek}, ``{Unified Channel Model for Mobile
  Radio Systems with Smart Antennas},'' \emph{IEE Proc. - Radar, Sonar and
  Navigation}, vol. 145, no.~1, pp. 32--41, Feb. 1998.

\bibitem{cost1}
A.~F. {Molisch}, H.~{Asplund}, R.~{Heddergott}, M.~{Steinbauer}, and
  T.~{Zwick}, ``{The COST259 Directional Channel Model-Part I: Overview and
  Methodology},'' \emph{IEEE Trans. on Commun.}, vol.~5, no.~12, pp.
  3421--3433, Dec. 2006.

\bibitem{cost2}
H.~{Asplund}, A.~A. {Glazunov}, A.~F. {Molisch}, K.~I. {Pedersen}, and
  M.~{Steinbauer}, ``{The COST 259 Directional Channel Model-Part II:
  Macrocells},'' \emph{IEEE Trans. on Wireless Commun.}, vol.~5, no.~12, pp.
  3434--3450, Dec. 2006.

\bibitem{costiracon}
C.~Gustafson, K.~Mahler, D.~Bolin, and F.~Tufvesson, ``{The COST IRACON
  Geometry-Based Stochastic Channel Model for Vehicle-to-Vehicle Communication
  in Intersections},'' \emph{IEEE Trans. on Veh. Technology}, vol.~69, no.~3,
  pp. 2365--2375, Mar. 2020.

\bibitem{bai22}
J.~G. Andrews, T.~Bai, M.~N. Kulkarni, A.~Alkhateeb, A.~K. Gupta, and R.~W.
  Heath, ``{Modeling and Analyzing Millimeter Wave Cellular Systems},''
  \emph{IEEE Trans. on Commun.}, vol.~65, no.~1, pp. 403--430, Jan. 2017.

\bibitem{corrblocking}
S.~Aditya, H.~S. Dhillon, A.~F. Molisch, and H.~M. Behairy, ``{A Tractable
  Analysis of the Blind Spot Probability in Localization Networks Under
  Correlated Blocking},'' \emph{IEEE Trans. on Wireless Commun.}, vol.~17,
  no.~12, pp. 8150--8164, Oct. 2018.

\bibitem{clone1}
C.~E. O’Lone, H.~S. Dhillon, and R.~M. Buehrer, ``{Characterizing the
  First-Arriving Multipath Component in 5G Millimeter Wave Networks: TOA, AOA,
  and Non-Line-of-Sight Bias},'' \emph{IEEE Trans. on Wireless Commun.},
  vol.~21, no.~3, pp. 1602--1620, Mar. 2022.

\bibitem{clone2}
------, ``{Single-Anchor Localizability in 5G Millimeter Wave Networks},''
  \emph{IEEE Wireless Commun. Letters}, vol.~9, no.~1, pp. 65--69, Jan. 2020.

\bibitem{pandey2023coverage}
K.~Pandey, A.~K. Pandey, A.~K. Gupta, and H.~S. Dhillon, ``{Coverage Analysis
  of a THz Cellular Network in the Presence of Scatterers},'' in \emph{Proc.,
  IEEE Intl. Conf. on Commun. (ICC)}, May 2023.

\bibitem{hyperbole}
S.~S. Mahmoud, Z.~M. Hussain, and P.~O'Shea, ``{A Space-time Model for Mobile
  Radio Channel with Hyperbolically Distributed Scatterers},'' \emph{IEEE
  Antennas Wirel. Propag. Lett.}, vol.~1, pp. 211--214, 2002.

\bibitem{ToA1}
A.~Y. Olenko, K.~T. Wong, and S.~A. Qasmi, ``{Distribution of the Uplink
  Multipaths’ Arrival Delay and Azimuth-elevation Arrival Angle because of
  ‘Bad Urban’ Scatterers Distributed Cylindrically above the Mobile},''
  \emph{Trans. on Emerging Telecommunications Technologies}, vol.~24, no.~2,
  pp. 113--132, 2013.

\bibitem{stoyan}
D.~Stoyan, W.~S. Kendall, S.~N. Chiu, and J.~Mecke, \emph{{Stochastic Geometry
  and its Applications}}.\hskip 1em plus 0.5em minus 0.4em\relax John Wiley \&
  Sons, 2013.

\bibitem{cluster}
H.~{Asplund}, A.~F. {Molisch}, M.~{Steinbauer}, and N.~B. {Mehta},
  ``{Clustering of Scatterers in Mobile Radio Channels - Evaluation and
  Modeling in the COST259 Directional Channel Model},'' in \emph{Proc., IEEE
  Intl. Conf. on Commun. (ICC)}, vol.~2, Apr. 2002, pp. 901--905.

\bibitem{goldsmith_pathloss}
A.~Goldsmith, ``{Path Loss and Shadowing},'' in \emph{Wireless
  Communications}.\hskip 1em plus 0.5em minus 0.4em\relax Cambridge University
  Press, 2005, p. 27–63.

\end{thebibliography}

\end{document}